# The interaction of He⁻ with fullerenes

Andreas Mauracher,[1] Matthias Daxner,[1] Stefan E. Huber,[1] Johannes Postler,[1] Michael Renzler,[1] Stephan Denifl,[1] Paul Scheier[1,*] and Andrew M. Ellis[2,*]

[1] Institut für Ionenphysik und Angewandte Physik, Universität Innsbruck, Technikerstr. 25, A-6020 Innsbruck, Austria

[2] Department of Chemistry, University of Leicester, University Road, Leicester, LE1 7RH, UK

Email: Paul.Scheier@uibk.ac.at; andrew.ellis@le.ac.uk

**Abstract**

The effects of interactions between He⁻ and clusters of fullerenes in helium nanodroplets are described. Electron transfer from He⁻ to $(C_{60})_n$ and $(C_{70})_n$ clusters results in the formation of the corresponding fullerene cluster dianions. This unusual double electron transfer appears to be concerted and is most likely guided by electron correlation between the two very weakly bound outer electrons in He⁻. We suggest a mechanism which involves long range electron transfer followed by the conversion of $He^+$ into $He_2^+$, where formation of the He-He bond in $He_2^+$ releases sufficient kinetic energy for the cation and the dianion to escape their Coulombic attraction. By analogy with the corresponding dications the observation of a size threshold for formation of both $(C_{60})_n^{2-}$ and $(C_{70})_n^{2-}$ of $n \geq 5$ is attributed to Coulomb explosion rather than a thermodynamic constraint. We also find that smaller dianions can be observed if water is added as a co-dopant. Other aspects of He⁻ chemistry that are explored include its role in the formation of multiply charged fullerene cluster cations and the sensitivity of cluster dianion formation on the incident electron energy.

**Introduction**

The helium monoanion, He⁻, was first detected in experiments in 1939.[1] He⁻ is an unusual ion because atomic helium in its $1s^2$ $^1S$ ground state has a negative electron affinity, on account of its compact closed-shell electronic structure. On the other hand metastable electronic excited states of helium are more polarizable than the ground state. The lowest of these metastable excited states, the $1s2s$ $^3S$ state, has a small positive electron affinity and can therefore attach an electron. Accurate calculations predict a binding energy of 77 meV for the resulting He⁻ ion in its $1s2s2p$ $^4P$ state,[2] a value confirmed by experimental measurements.[2,3] However, He⁻ is metastable with respect to autodetachment since it is embedded in a continuum of states formed from the ground state of neutral atomic helium and a free electron. The lifetimes of the spin-orbit components of the $1s2s2p$ $^4P$ state have been subjected to various measurements.[4-7] The most recent and precise measurement for the lowest and longest lived spin-orbit component, the $J = 5/2$ state, yields a value of 359.0 ± 0.7 μs.[8] The other spin-orbit components, corresponding to $J = 3/2$ and $1/2$, have even shorter lifetimes on the order of 10 μs.[6,7]

Given the short lifetime of He⁻ it is challenging to explore the physics and particularly the chemistry of this ion in the gas phase. However, the recent discovery of He⁻ in superfluid helium nanodroplets has opened up a new route for exploring the properties of this unusual anion.[9] Mauracher *et al.* showed that He⁻ ions can be formed by electron impact in sufficiently large helium droplets. The formation of He⁻ is a resonant process with maximum production at an electron energy of 22 eV. This resonance occurs because the formation of He⁻ first requires production of a metastable excited state of neutral He capable of binding an additional electron, such as the lowest $^3S$ state mentioned above. The lowest $^3S$ state, which we designate by the shorthand notation as He*, is at an energy 19.8 eV above the ground electronic state of the helium atom in the gas phase. However in a helium droplet additional energy is required for the electron to penetrate inside the liquid and form a cavity ('bubble') to accommodate the He*. The resulting slow electron from the inelastic scattering of the original incoming electron can then attach to the He* to make He¯. By analogy with an electron it is expected that He¯ will exist within a bubble inside a helium nanodroplet because of significant repulsive interactions with its immediate neighbor atoms.

In addition to He¯, $He_2^-$ is also formed by a resonant electron attachment process in helium droplets but is observed to be almost two orders of magnitude less abundant than He¯.[9] Furthermore, through the addition of a dopant to the droplet, $SF_6$, Mauracher *et al.* were able to show that He¯ is highly mobile and can seek out and transfer its negative charge to the dopant.[9] On the other hand, the diatomic helium anion is strongly heliophobic and presumably resides at or near the surface of the helium droplet. Consequently, any anion chemistry with dopants is expected to be dominated by the highly mobile He¯ ion.

With the ability to produce He¯ in close proximity with a dopant molecule or cluster inside a liquid helium droplet, it now becomes possible to explore the chemistry of this unusual anion. This can be seen rather clearly if we assume that the He¯ bubble can travel at the Landau velocity for superfluid $^4$He (*ca.* 60 m s$^{-1}$).[10] In this case the He¯ can traverse a droplet of a size on the order of $10^5$ helium atoms in < 1 ns, which is many orders of magnitude below the measured autodetachment lifetime of He¯ in its 1s2s2p $^4$P state. Recently, we published a short report on electron transfer from He¯ to fullerene clusters, $(C_{60})_n$ and $(C_{70})_n$, in helium nanodroplets.[11] A remarkable finding was that, in addition to monoanions produced by direct electron attachment, dianions were also formed. This was the first report of dianions produced inside helium droplets and the dependence of the dianion signal on the electron energy was found to possess the same resonance behavior as for He¯ formation, showing that the dianions were produced by two-electron transfer from He¯. It seems likely that the two-electron transfer from He¯ is facilitated by having two weakly bound electrons orbiting what is essentially a He$^+$ core. Under these conditions one expects that electron correlation between the two outermost electrons will be a major contributor to the 77 meV stabilization provided by adding an electron to He*. It therefore seems reasonable to think that electron correlation is also important in driving the two-electron transfer from He$^-$ to the fullerenes.

In this paper we explore several facets of the interaction between He$^-$ and fullerene clusters in more detail, as well as reporting several new findings. We discuss the two-electron transfer to helium and show that the minimum fullerene cluster size needed to see dianions is a consequence of rapid Coulomb explosion for smaller dianions. We also explore the role of He$^-$ in the formation of multiply charged cations. Finally, we discuss the production of He$^-$ at energies well above 22 eV.

**Experimental**

Full details of the apparatus have been given previously[12] and so only a brief account is provided here. Helium nanodroplets were produced by expanding highly pure (99.9999%) gaseous helium at high pressure (21–23 bars) and a controlled temperature (8.8–9.6 K) through a 5 μm pinhole into a vacuum. Expansion conditions in the current work were chosen to yield helium nanodroplets with a mean size exceeding $1.8 \times 10^5$ helium atoms, the minimum required to form He$^-$.[9] The expansion was skimmed to form a collimated droplet beam and this was then passed through a heated pick-up cell containing either $C_{60}$ or $C_{70}$ (SES Research, purity 99.95% and 99%, respectively). After the pick-up of fullerene molecules from the vapour produced in the heated pick-up cell the droplets passed through a second skimmer to enter another differentially pumped chamber, where they were exposed to an electron beam of variable energy (0-150 eV). Any ions produced were then extracted into a high resolution and high repetition rate reflectron time-of-flight mass spectrometer (Tofwerk).

**Results and Discussion**

1.  *Fullerene cluster dianion size distributions*

Figure 1 shows a negative ion mass spectrum for helium droplets containing clusters of $C_{60}$ recorded at an incident electron energy of 22 eV, *i.e.* at the peak resonance for $He^-$ production. The dominant anions seen are the monoanions, $(C_{60})_n^-$. At much lower intensities (note the logarithmic vertical scale) signals from the dianions $(C_{60})_n^{2-}$ can also be seen. As shown elsewhere, we attribute the formation of the dianions to the transfer of two electrons from $He^-$ to the neutral fullerene cluster.[11] This electron transfer could occur stepwise from two distinct $He^-$ ions formed within a helium droplet but transfer of the second electron would then be complicated by electrostatic repulsion between the two negatively charged species. Experiments by Schweikhard and co-workers have shown that in order to attach an electron to a monoanion a minimum electron kinetic energy of typically several eV is required in order to overcome the Coulomb barrier.[13] Assuming that $He^-$ travels no faster than the Landau velocity in superfluid helium,[10] which is near to 60 m s$^{-1}$, then the kinetic energy of a $He^-$ ion will be <1 meV, which is far too low to exceed the expected Coulomb barrier.

To provide an alternative mechanism for two-electron transfer, we recently proposed the process summarized (simplistically) in the reaction below,

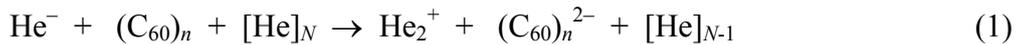

$$He^- \ + \ (C_{60})_n \ + \ [He]_N \ \rightarrow \ He_2^+ \ + \ (C_{60})_n^{2-} \ + \ [He]_{N-1} \qquad (1)$$

where $[He]_N$ represents a helium droplet composed of N helium atoms. The suggestion here is that both electrons are transferred to the fullerene cluster from a single $He^-$ ion simultaneously, perhaps driven by the strong electron correlation between the two outermost electrons in He⁻. This assumption bypasses the problem of electron-electron repulsion inherent in a two-step electron transfer. Although it has not been reported previously for a two-electron transfer, at least as far as we are aware, a simultaneous two-electron transfer is possible at long range by the well-known harpoon mechanism.[14] This model identifies the threshold distance at which electron transfer can take place as the crossing point between the ion-induced dipole potential energy surface between the He⁻ and the fullerene cluster and the potential energy surface derived from the long range Coulombic interaction between the fullerene dianion and He⁺. With this model we can calculate the electron transfer distance, $r_c$, from the relationship:

$$\frac{e^2}{2\pi\varepsilon_0 r_c} = (IP)_d - (EA)_d \qquad (2)$$

Here $(IP)_d$ is the double ionization energy of He⁻, which from a combination of the first ionization energy of helium (24.6 eV), the excitation energy of He* (19.82 eV) and the bidning energy of the outermost electron in He⁻ (0.08 eV) we arrive at a value of 4.69 eV. The other quantity needed, the double electron affinity of the fullerene cluster, $(EA)_d$, is more uncertain but values derived from DFT calculations were reported in ref. 11 for small $(C_{60})_n$ clusters. As an illustration the calculated value for $(C_{60})_4$ is $(EA)_d = 4.5$ eV. From equation (2) we then determine $r_c \approx 150$ Å. We do not attach much precision to this large value because of the approximations involved but the calculation does at least show that long range two-electron transfer from He⁻ to a fullerene cluster is plausible from an energetic standpoint. Having transferred the electrons at long range, the resulting He⁺ can then combine with a nearby He atom to form $He_2^+$, which releases 2.5 eV of energy.[15-17] It is through this energy release that we account for the separation of the cation and the dianion. If all of the excess energy is deposited into kinetic energy then we estimate that the two ions can escape Coulombic attraction even if electron transfer occurs at a distance as short as 12 Å. The same general process is assumed to operate for dianion formation in clusters of $C_{70}$.

We now turn to the observed size distributions of the $(C_{60})_n^{2-}$ and $(C_{70})_n^{2-}$ cluster ions, and in particular the appearance thresholds. It is noteworthy that in both cases the smallest observable dianion is found at $n = 5$. For $(C_{60})_n^{2-}$ this is seen in the plot of dianion abundances shown in the inset of Figure 1. There are two possible explanations for this specific onset: (1) smaller dianion clusters are thermodynamically inaccessible and so do not form or (2) the smaller dianion clusters are formed but are unstable with respect to Coulomb explosion. Using DFT calculations to calculate double electron affinities for small $(C_{60})_n$ clusters we previously predicted that for $n < 4$ formation of the dianion is thermodynamically forbidden in a two-electron transfer from $He^-$.[11] Given the margin of uncertainty in these calculations it is plausible that the true thermodynamic threshold for dianion formation begins at $n = 5$, which would account for the experimental appearance threshold. However, work by Zettergren and co-workers on dications of $C_{60}$ clusters, $(C_{60})_n^{2+}$, has found the smallest observed cluster ion is $(C_{60})_5^{2+}$.[18] This equivalence in the smallest observable cluster size for both dianions and dications suggests that Coulomb explosion is the primary factor determining the minimum observed size of the cluster dianions. Support for this is also provided by comparing the experimental monoanion cluster size distribution to that of the dianion, which are both shown in the inset of Figure 1. The monoanion size distribution is shifted to lower cluster sizes than that for the dianions, presumably because Coulomb explosion of small dianion clusters enhances the signal levels for small monoanions.

We have also found that addition of a second dopant to the fullerene clusters can change the minimum size distribution for the fullerene cluster dianion. For example, addition of a single water molecule now stabilizes $(C_{60})_4^{2-}$, as seen in the mass spectrum shown in Figure 2(a). As another example, addition of 11 water molecules enables $(C_{60})_3^{2-}$ to be observed (see Figure 2(b)). The precise role of the water in this stabilization process remains to be established but presumably it is a form of solvent stabilization of the excess negative charge which helps to reduce the Coulombic repulsion and therefore the probability of Coulomb explosion.

2. *Fullerene monomer dianions*

In addition to the $(C_{60})_n^{2-}$ and $(C_{70})_n^{2-}$ cluster ions, we have also managed to identify exceptionally weak signals from the monomer dianions, $C_{60}^{2-}$ and $C_{70}^{2-}$. The $C_{60}^{2-}$ ion is barely detectable whereas $C_{70}^{2-}$ is more clearly observable and a mass spectrum derived from this species is shown in Figure 3. Neither $C_{60}^{2-}$ nor $C_{70}^{2-}$ can undergo Coulomb explosion because of the strong covalent bonds holding the fullerenes together. Instead the decay of these ions can only occur through autodetachment of one of the electrons. In fact it is known from work performed in the gas phase that cold $C_{70}^{2-}$ ions are stable against autodetachment and are essentially long-lived ions because both the first and second electron affinities of $C_{70}$ are positive.[19] In contrast $C_{60}^{2-}$ is metastable because $C_{60}$ has a negative second electron affinity.[20] Nevertheless, $C_{60}^{2-}$ has been found to survive for relatively long periods, at least on the millsecond timescale at low temperatures,[21] because of a Coulomb barrier that inhibits the loss of an electron.[22]

A key question to answer is how these dianions form in helium droplets? The production of $C_{60}^{2-}$ and $C_{70}^{2-}$ by two-electron transfer from He⁻ to the neutral fullerene monomers is thermodynamically forbidden. For example, an additional 1.91 eV is needed to reach the thermodynamic threshold energy to produce $C_{70}^{2-}$, with the formation of $C_{60}^{2-}$ being even less favorable. However, these barriers will be reduced if the He⁻ that collides with $C_{60}$ or $C_{70}$ is in a higher electronic state. We defer discussion of such states until a later section, but we note here that a state lying 1 eV above the lowest bound state of He⁻ is identified later which would substantially reduce the energy barrier. Such reactions would still be energetically unfavorable but this would explain the very low yield of $C_{60}^{2-}$ and $C_{70}^{2-}$ seen in the experiments.

Although the $C_{70}^{2-}$ signal is extremely weak, it has proved possible to record the ion yield as a function of incident electron energy. The data obtained, which are extremely noisy because of the low signal level, are shown in the lower panel in Figure 3. The solid blue line is a fit to the experimental data, and although there are considerable uncertainties, there is clear evidence of a peak in production at an energy consistent with formation of $C_{70}^{2-}$ by electron transfer from He⁻. Furthermore, although the data are noisy, this peak is found at 23 eV, approximately 1 eV above the peak for He⁻ (1s2s2p) production and therefore consistent with reaction by a higher lying electronic state of He⁻.

For $C_{60}^{2-}$, the signal was even weaker than for $C_{70}^{2-}$. In fact, it was found that we could only detect $C_{60}^{2-}$ dianions when water was added to the second pick-up cell, where the role of the water is unknown. There is no report in the literature of $C_{60}^{2-}$ being made by direct electron attachment to the anion, although it has been made by colliding Na atoms with $C_{60}^{-}$ at very high collision energies.[23] The exceptionally weak $C_{60}^{2-}$ signal may simply reflect the fact that the energetics for electron transfer from He⁻ to $C_{60}$ is even less favorable than for $C_{70}$.

3. *Formation of multiply charged cations*

Multiply charged cationic fullerene clusters, $(C_{60})_n^{q+}$, as well as dianions, can also be seen in the mass spectra. In the case of dications our findings agree with those of Zettergren *et al.*,[18] with the smallest observed $(C_{60})_n^{2+}$ ion corresponding to *n* = 5. However, more highly charged fullerene cluster cations can also be observed. For example, Figure 4 shows spectra with peaks from $(C_{60})_n^{3+}$ and $(C_{60})_n^{4+}$ ions, which are assigned unambiguously on account of their absolute masses, and the fact that a series of such ions with different values of $n$ has been seen for each charge, $q$.

Although the signals are weak, particularly for the triply and quadruply charged cations, some information about magic number ions and also about the appearance thresholds has been extracted. Our observations are similar to those reported previously in gas phase studies.[18,24] This earlier work identified magic number ions, *i.e.* ions with anomalously high abundance relative to the neighboring cluster sizes, at $n = 13$ for $q = 2$ and at 13, 19 and 23 for $q = 3$. We see magic number peaks at $n = 13$ for both $q = 2$ and 3 and we also see magic number peaks at $n = 19$ for both types of ions. Exact agreement with the gas phase phase work is not maintained for larger $q = 2$ and 3 cluster ions, as well as for $q = 4$. However, the fact that we see magic numbers at all suggests that, as in the gas phase work, evaporation processes must be occurring in the helium droplets. In other words, rapid quenching in the helium droplets does not occur following ionization, otherwise no magic number ions would be observed.

The appearance sizes for the cluster ions in the current study are also very similar to earlier gas phase work.[18,24] As identified above, only ions with $n \geq 5$ are stable against Coulomb explosion for the dications. As expected the threshold appearance sizes are larger for more highly charged clusters, as summarized in Table 1. The slightly larger threshold sizes observed in this work for $q = 3$ and 4 presumably derive from the very weak signals we observe near to threshold, which makes it difficult to ascertain the true threshold size.

How can these multiply charged fullerene cluster ions be formed in helium nanodroplets? The formation of dopant cations in helium nanodroplets is normally attributed to initial formation of $He^+$ as an electron strikes the droplet. The positive charge can then hop from atom to atom until finding the dopant and transferring the charge.[25-28] However, this mechanism is an unlikely route for generating multiply charged cations because electrostatic repulsion would inhibit this process. An alternative possibility is that He* brings about ionization through the Penning process[29], i.e.

$$\text{He}^* + (\text{C}_{60})_n \rightarrow \text{He} + (\text{C}_{60})_n^+ + e^- \quad (3)$$

and subsequently through further reactions after additional electron strikes, *i.e.*

$$\text{He}^* + (\text{C}_{60})_n^{z+} \rightarrow \text{He} + (\text{C}_{60})_n^{(z+1)+} + e^- \quad (4)$$

In the case of $C_{60}$ the He* is capable of providing enough energy to remove an electron from $C_{60}^+$ (*ca.* 11.4 eV[30]) and $C_{60}^{2+}$ ((*ca.* 17 eV[30]) but not enough for $C_{60}^{3+}$ (*ca.* 27 eV[30]). Nevertheless, these ionization energies would be expected to decline markedly for $(C_{60})_n^{z+}$ clusters, particularly when *n* is rather large, as is the case for the examples shown in Figure 4. Formation of cluster ions as highly charged as $(C_{60})_n^{4+}$ clusters is therefore energetically plausible via Penning ionization.

A problem with the He* mechanism is that He* is known be heliophobic and therefore will reside at or close to the helium droplet surface.[17] In contrast the fullerene cluster will be fully solvated by the helium and will locate itself somewhere near the center of the droplet. Consequently, in order for Penning ionization to take place, either the He* must migrate inwards or it must be formed by electron impact somewhere near the fullerene cluster, whereby it will then be drawn towards the fullerene by the polarizability of the latter.

While Penning ionization might be the source of the triply and quadruply charged ions in Figure 4, there is another and perhaps more likely mechanism involving $He^-$ as the ionizing agent. Since the data presented in Figure 4 were recorded at an electron energy of 85 eV which is well above the optimum energy for forming $He^-$, the formation of $He^-$ is still possible through multiple electron-helium collisions, which can deliver a low energy electron in the vicinity of He*. An alternative route to He¯ is via combination of two low energy electrons with He⁺ (see section titled "Effect of the electronic state of He¯"). In contrast to He*, He¯ is known to be heliophilic and highly mobile within helium droplets.[19] In particular, once a positively charged fullerene cluster is formed, the strong Coulombic attraction between He¯ and the cation will mean that the former is likely to be drawn directly towards the latter. Thus, through a sequence of He¯ production events, multiply charged cations can be generated by the reaction shown below, a form of Penning ionization involving an anion as the energy source,

$$\text{He}^- + (\text{C}_{60})_n^{z+} \rightarrow \text{He} + (\text{C}_{60})_n^{(z+1)+} + 2\text{e}^- \qquad (5)$$

The sequential production of He¯ in helium droplets by consecutive electron strikes will be favored in large droplets because they possess large collision cross sections. Thus the observation of multiply charged fullerene cluster cations of relatively large size in Figure 4 may be no accident, since these clusters will most likely be formed in large helium droplets.

In future work, it would be interesting to explore in more detail the mechanism by which these multiple charge cluster ions form. Useful information might be obtained from a study of the dependence of the ion signal on the incident electron energy and on the electron current.

4. *Effect of the electronic state of He¯*

The efficiency of electron donation from He¯ is sensitive to the incident electron energy. Before considering the impact on fullerene dianion formation, we first highlight findings for electron donation to SF₆ in helium nanodroplets. In the case of SF₆ only monoanion formation is induced by electron transfer from He⁻, presumably because formation of dianions is thermodynamically forbidden (there may also be other factors that would prevent observation of dianions, such as Coulomb explosion). The primary anionic product detected in the gas phase by mass spectrometry is the fragment $SF_5^-$. Figure 5 shows how the He⁻ ion yield is affected by addition of $SF_6$. Some preliminary data of this type were briefly presented in ref. 9 but without explanation.

In the absence of a dopant the He⁻ yield curve is strongly peaked at 22 eV but is clearly asymmetric, with a tail to higher electron energies. As shown previously,[9] this ion yield curve can be modelled reasonably well using three Gaussian functions, with the highest weighting given to a Gaussian centered at 22 eV and two much weaker peaks having maxima at *ca.* 23 and 25 eV. The 25 eV peak is markedly broader than the two lower energy peaks, as becomes even more obvious when $SF_6$ is added, as we discuss shortly. The interpretation of the 22 eV peak is, as discussed earlier, that it corresponds to formation of He⁻ in its lowest metastable state, the 1s2s2p $^4$P state. The 23 eV peak is consistent with production of He⁻ in a higher lying excited electronic. However, we note that in previous experimental work only two bound states of He⁻ have been reported, namely the 1s2s2p $^4$P and the 2p³ $^4$S states.[31,32] The 2p³ $^4$S state is a much higher energy state, with the best available calculations predicting that this state lies ca. 39.6 eV above the 1s2s2p $^4$P state. We therefore conclude that the 2p³ $^4$S state plays no role in our experiments and therefore the resonance in the He⁻ state at 23 eV is attributed to an excited state of He⁻ that has not been identified previously as a bound state. In a recent theoretical study Huber and Mauracher suggested that the 1s2p² $^4$P state of He⁻ might be responsible,[17] since it is expected to lie roughly 1 eV above the 1s2s2p $^4$P state. This assignment seems plausible but it is surprising that the 1s2p² $^4$P state of He⁻ has not been reported in earlier gas phase experiments.

The source of the highest energy resonance, the peak centered near to 25 eV, also needs explaining. One could attempt to identify other excited electronic states of He⁻ to account for this feature but the challenge then would be to explain the long tail extending to much higher electron kinetic energies (> 30 eV). Huber and Mauracher have suggested that this peak might have a very different source, arising from recombination of He⁺ with two electrons, since the energy of the peak maximum lies is only marginally above the ionization threshold of atomic helium in its ground state,[17] i.e.,

$$e^- + He(^1S) \rightarrow 2e^- + He^+(^2S) \rightarrow He^-(^4P) \qquad (6)$$

Attachment of both electrons to He⁺ can only occur if these two electrons have near zero kinetics energies. Inelastic scattering by the surrounding helium could deliver these low energy electrons. However, the ability of the helium to both quench and confine both electrons is expected to decline with increasing excess energy, which would explain the long tail to higher electron kinetic energies.

When SF₆ is added, a dramatic change in the He⁻ yield curve is seen (Figure 5). Most significantly there is a large suppression of the total He⁻ signal, which is attributed to electron transfer to the SF₆. However, it is also found that the He⁻ produced at the two lower energy resonances (22 and 23 eV) is more strongly affected by the addition of SF₆ than that at high energy resonance (25 eV). If the 25 eV peak was derived from reaction (6) above, this different behavior would not be expected since it would result in He⁻ being formed in the same state as for the 22 eV resonance. We think that a more likely explanation is that the He⁻ is formed in one or more higher energy electronic states at ≥25 eV. Presumably, these will be very diffuse Rydberg-like states which will be more reactive than the lower states of He⁻ because of the additional internal energy. Like the 1s2p² ⁴P state assumed to be responsible for the 23 eV resonance, which was discussed earlier, this suggestion requires production of additional excited states of He⁻ which have not been observed previously in gas phase experiments, most likely because they undergo rapid autodetachment. However, it is possible that the surrounding helium atoms in a liquid helium droplet impede the autodetachment process and allow the He⁻ to survive for long enough to transfer one or two electrons to a dopant.

We turn now to the fullerenes to see if these shed any further light on this issue. Figure 6 shows how the relative intensity (as determined by peak area) of the 25 eV peak versus the combined intensities of the 22 and 23 eV peaks varies with the size of the detected fullerene cluster dianion, $(C_{60})_n^{2-}$. To extract the relative peak areas the 22 and 23 eV peaks were represented by Gaussian functions whereas the 25 eV peak was expressed as the combination of a Gaussian and a log-normal function, where the latter accounts for the long tail to high energies. It seems likely that this long tail arises from electrons whose kinetic energy has been quenched by collisions within the helium droplet. Figure 6 shows a very marked increase in the 25 eV peak versus the combined 22 + 23 eV peaks with cluster size, n. This observation matches the $SF_6$ findings.

Finally, we show in Figure 7 the ion yield curve for He⁻ in the absence of any dopant for three different nozzle temperatures. At the highest temperature, which corresponds to the smallest mean helium droplet size ($4 \times 10^4$ helium atoms),[33] the 22 eV peak is prominent and there is a modest asymmetry which could account for a small contribution at 23 eV. However, there is clearly no long tail at ≥25 eV peak for these small droplets. A contribution from this tail becomes discernible for the larger droplets and is particularly noticeable at a nozzle temperature of 8.2 K (mean droplet size $5 \times 10^6$ helium atoms).[33] This dependence on the helium droplet size can be explained readily by assuming that the 25 eV peak results from a very diffuse Rydberg state of He⁻, as suggested earlier. One can even imagine a diffuse

Rydberg state where the outer electron orbits outside of the helium droplet. We think that this is unlikely because it does not explain the droplet size dependence reflected in Figure 7. Instead, we suggest that the Rydberg state is located within the droplet and only those droplets that are sufficiently large will be able to accommodate the large bubble needed to stabilize this diffuse form of He⁻. Presumably, this excited form of He⁻ can eventually relax to the lowest state of He⁻, leading to its ejection from the helium droplet and allowing detection by mass spectrometry.

**Conclusions**

The role of He⁻ in various observations has been reported. We have shown that it is possible for He⁻ to act as a two-electron donor, where the receptor species is a fullerene cluster. This two electron transfer can occur at long range and may happen simultaneously, perhaps driven by strong electron correlation between the two outer electrons in He⁻. The pure fullerene cluster dianions $(C_{60})_n^{2-}$ and $(C_{70})_n^{2-}$ are only seen for $n \geq 5$, the same threshold as seen for the corresponding doubly charged cations and which therefore implies that it is Coulomb explosion that dictates the anion threshold. We have also shown that the dianion threshold can be changed by adding water molecules, with even $(C_{60})_3^{2-}$ being stabilized when sufficient water is added.

He⁻ has also been implicated as the primary reagent in forming multiply charged cations in helium droplets. Although one normally thinks of charge transfer from He⁺ as being the major source of dopant cations in helium droplets, He⁺ would be unsuitable for making multiply charged cations because of Coulombic repulsion between a second He⁺ and an existing dopant cation. However, there are no such limitations for the highly mobile He⁻ ion and this ion possesses ample excess electronic energy (*ca.* 20 eV) to ionize many species, including the singly and indeed multiply charged fullerene cluster cations explored in this study.

Finally, we have reported the influence of the incident electron kinetic energy on the formation of He¯ and its interaction with both $SF_6$ and fullerenes. We see clear evidence of the role of two bound states of He¯, the $1s2s2p$ $^4P$ and $1s2p^2$ $^4P$ states, as reflected by resonances in the He¯ ion yield curves at 22 and 23 eV. Surprisingly, there is no evidence in the literature to suggest that the $1s2p^2$ $^4P$ state is a bound state of He¯, so our assignment must be considered as tentative. A higher energy feature in the ion yield curve is also seen which peaks above 25 eV and which becomes prominent in the presence of a dopant in the helium droplet. We propose that the 25 eV peak results from the production of one or more even higher lying electronic states of He¯, none of which have been observed in previous experiments. These higher energy states are expected to be more reactive with dopants than the lower energy states of He¯, accounting for the enhanced reactivity seen in anion product channels for electron energies near to 25 eV.

**Acknowledgements**

This work was given financial support by the Austrian Science Fund (FWF)Wien (P23657, P26635, P24443, I1015, and I978). We also thank the referees for constructive suggestions for improving the manuscript.

**Table 1**  Observed appearance sizes for $(C_{60})_n^{q+}$

| Charge, $q$ | This work | Ref. 21 |
|:---:|:---:|:---:|
| +2 | 5 | 5 |
| +3 | 11 | 10 |
| +4 | 25 | 21 |

**Figure captions**

1. Anion survey mass spectrum. The dominant peaks arise from $(C_{60})_n^-$. At much lower abundance $(C_{60})_n^{2-}$ dianions can also be seen and are marked in red. The inset compares the observed size distributions of the monoanions and dianions.

2. Sections of the mass spectra showing stabilization of (a) $(C_{60})_4^{2-}$ by one water molecule and (b) $(C_{60})_3^{2-}$ by 11 water molecules.

3. Mass spectra showing peaks from (a) $C_{70}^{2-}$ and (b) $C_{70}^-$. Both ions show identical isotope features, as expected for ions that differ only in charge. Panel (c) shows the anion efficiency data for $C_{70}^{2-}$ (blue symbols) and $(C_{70})_5^{2-}$ (red solid line). The solid blue line is a strongly filtered fit to the experimental data points obtained for $C_{70}^{2-}$.

4. Cationic mass spectrum recorded at an electron energy of 85 eV and an electron filament current of 82 μA. Illustrative signals from clusters with charges of +3 and +4 are shown in the inset.

5. Dependence of the He⁻ signal on electron energy in the absence and in the presence of $SF_6$ dopant. Note that the ion yield curve for the non-doped droplets has been divided by a factor of 50 to fit onto the same scale as the plots generated with dopant added.

6. Variation of low energy (22+23 eV) and high energy (25 eV) components of the $(C_{60})_n^{2-}$ signal as a function of $n$.

7. Effect of droplet size on the He⁻ yield curve. The nozzle temperature employed and the mean helium droplet size (N helium atoms) are shown in the upper right.

.

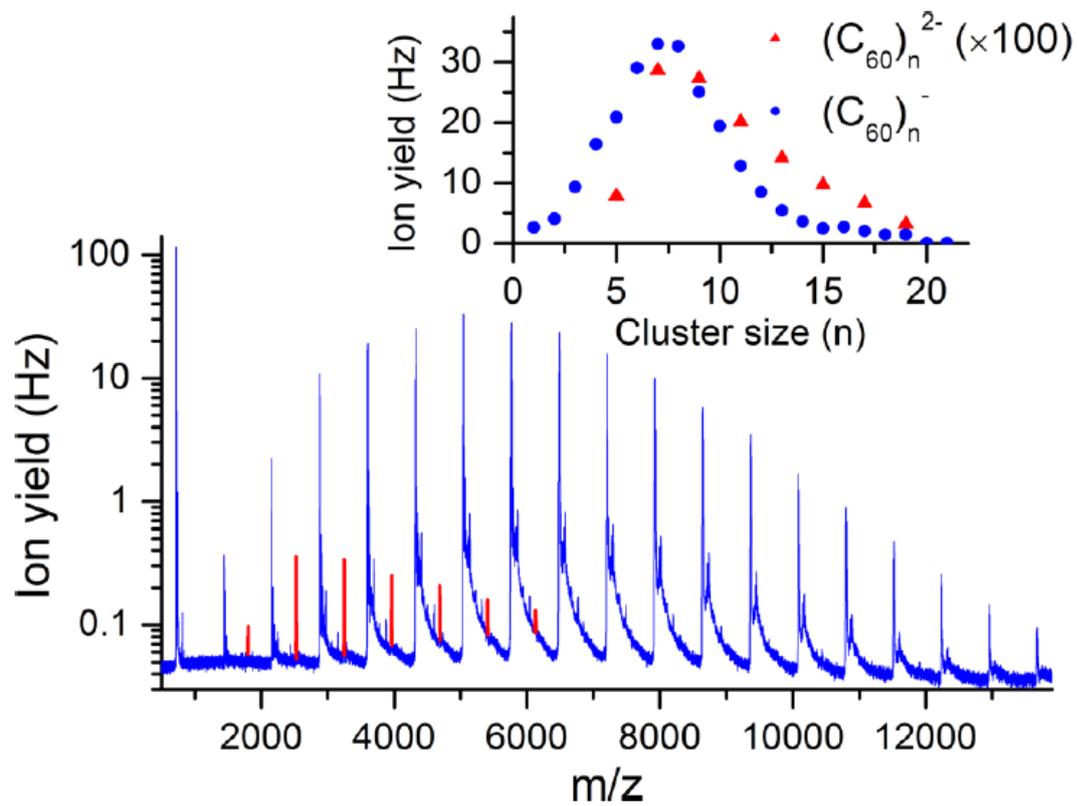

Figure 1

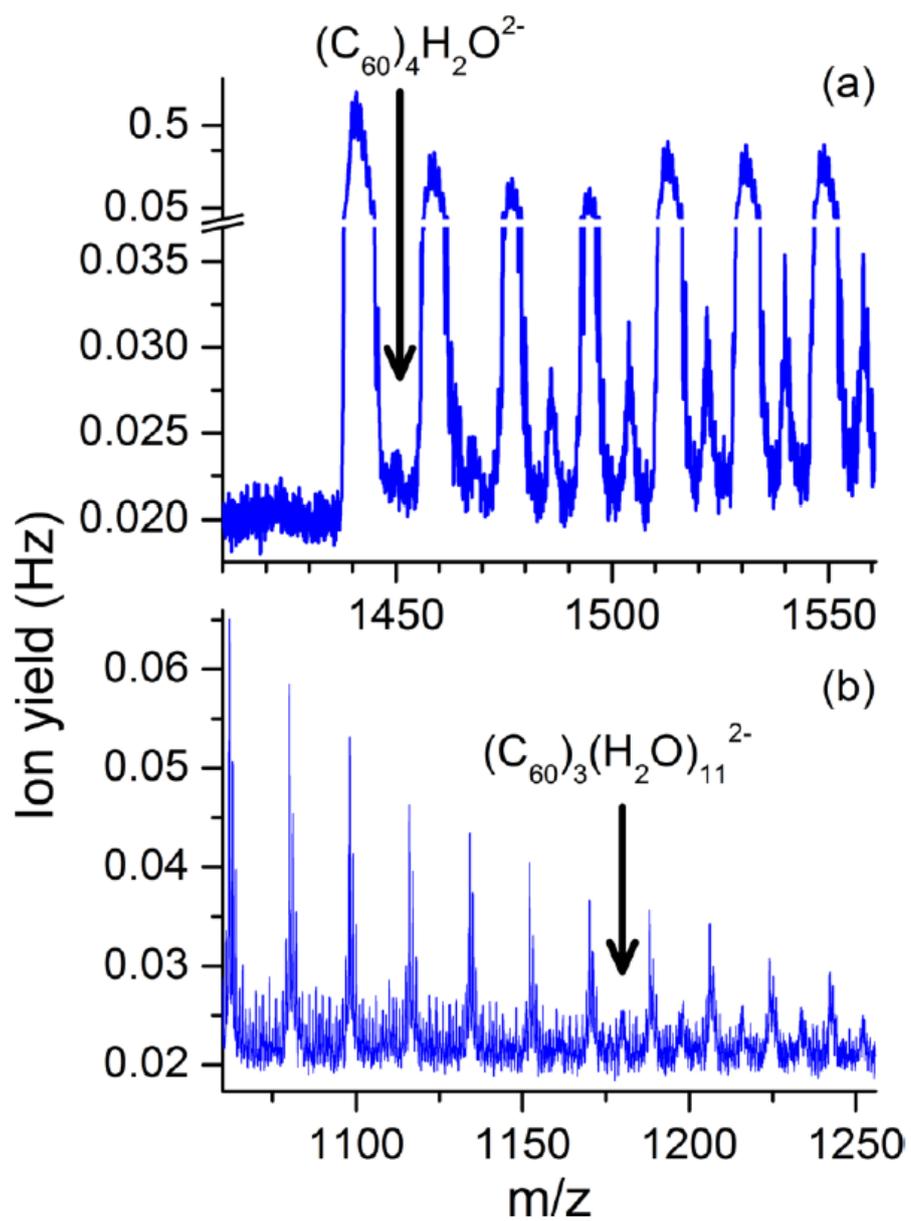

Figure 2

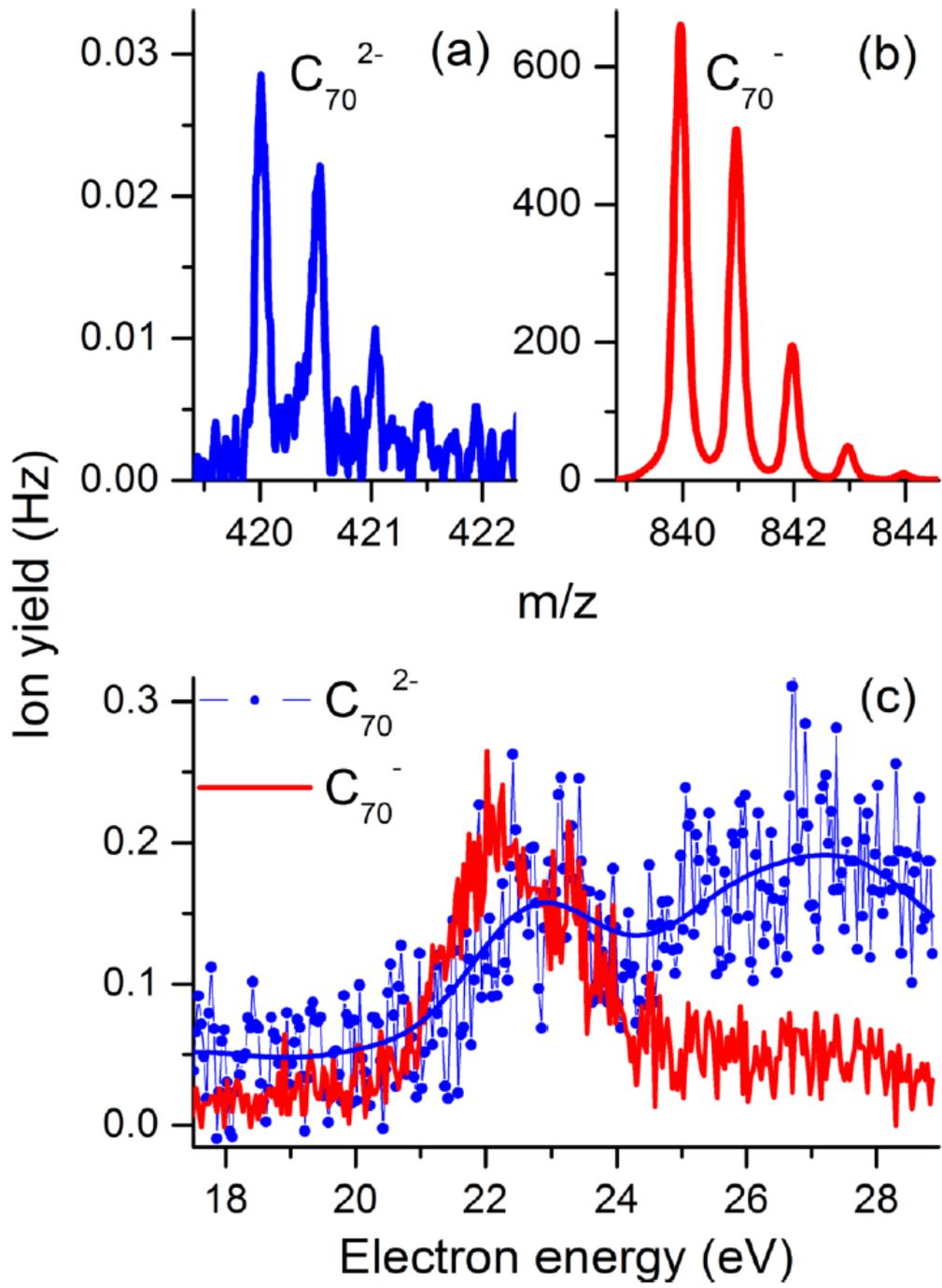

Figure 3

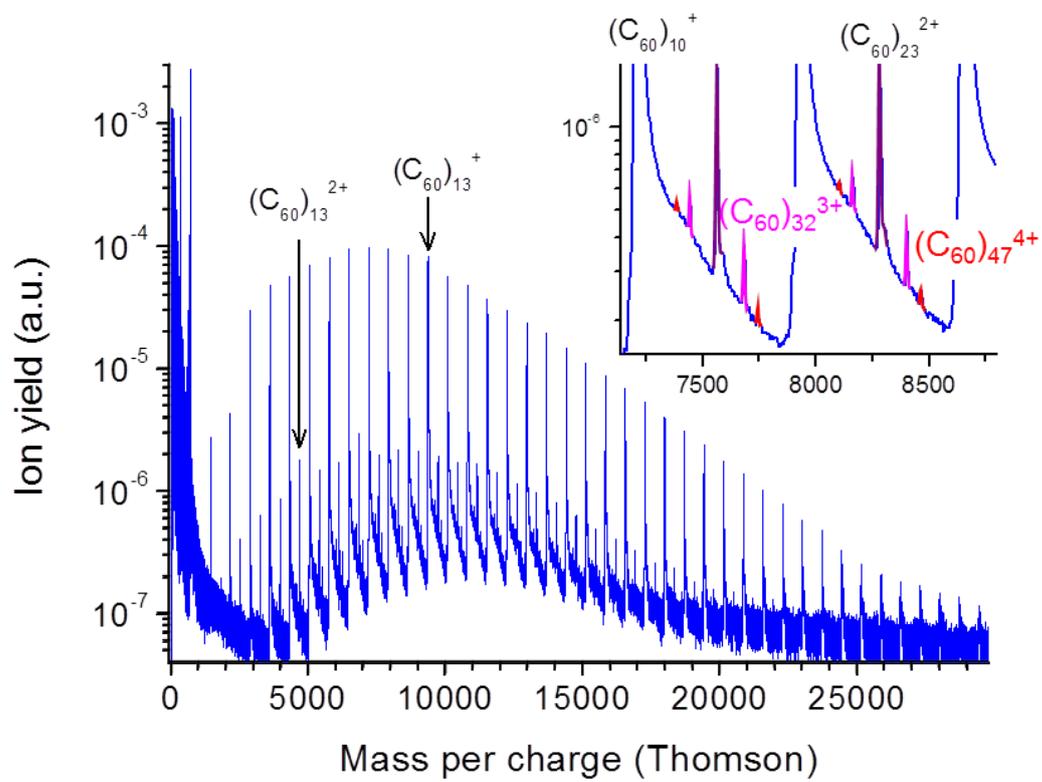

Figure 4

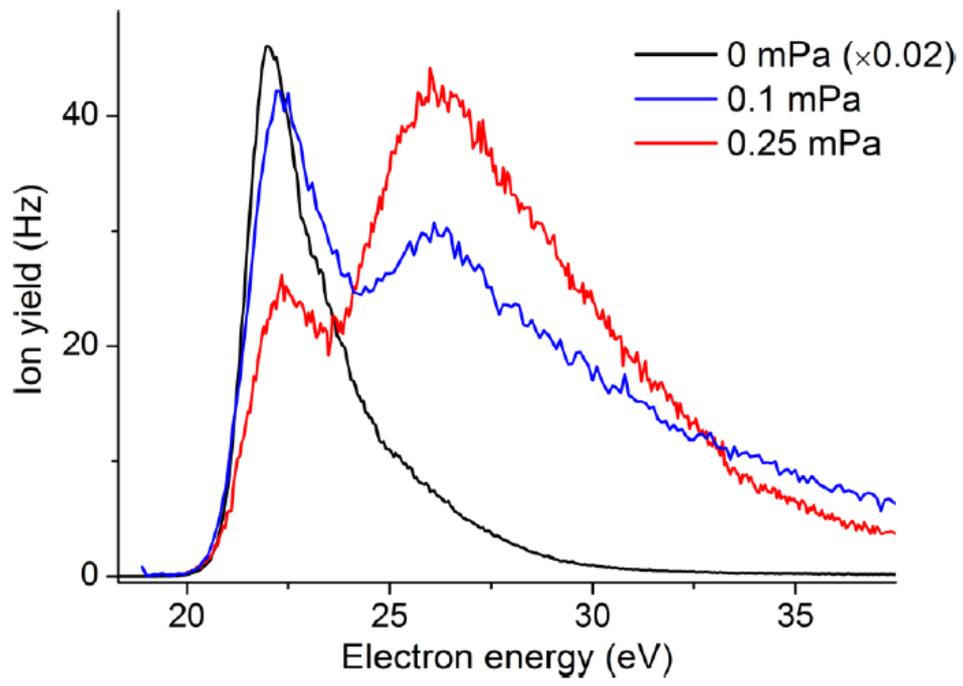

Figure 5

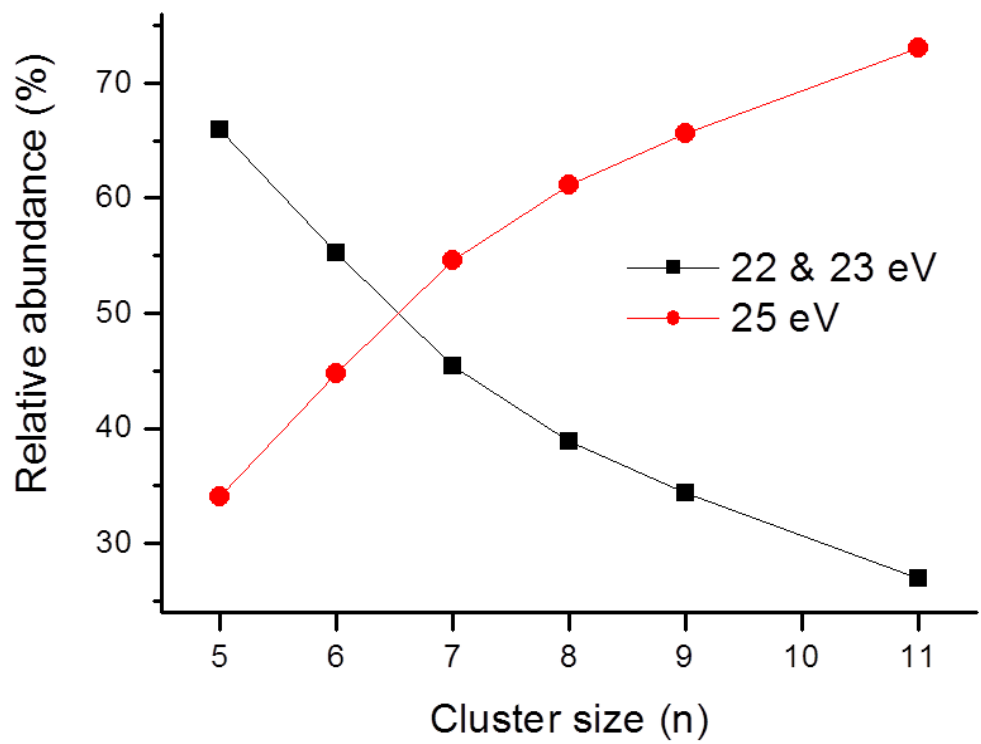

Figure 6

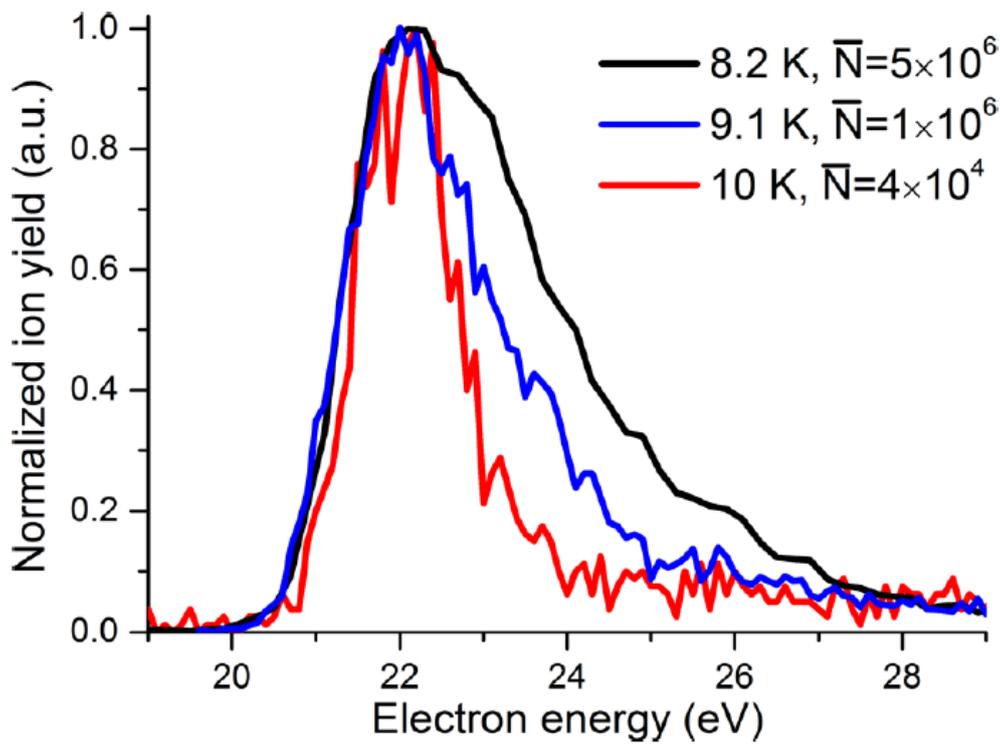

Figure 7